\documentclass{article}

\usepackage{amsmath,amssymb,amsthm}
\usepackage{esdiff}
\usepackage{color}
\usepackage{graphicx}
\usepackage{subfigure}

\title{Decreasing flow uncertainty in Bayesian inverse problems through Lagrangian drifter control}
\author{D. McDougall\footnotemark[2]
\and C.~K.~R.~T. Jones\footnotemark[3]}

\newcommand{\grad}{\ensuremath\nabla}

\begin{document}
\maketitle
\newcommand{\slugmaster}{%
\slugger{juq}{xxxx}{xx}{x}{x--x}}%slugger should be set to juq, siads, sifin, or siims

\renewcommand{\thefootnote}{\fnsymbol{footnote}}

\footnotetext[2]{Institute for Computational Engineering and Sciences, University of Texas at Austin, TX, USA}
\footnotetext[3]{Mathematics Department, University of North Carolina at Chapel Hill, NC, USA}

\renewcommand{\thefootnote}{\arabic{footnote}}

\begin{abstract}
Commonplace in oceanography is the collection of ocean drifter positions.
Ocean drifters are devices that sit on the surface of the ocean and move
with the flow, transmitting their position via GPS to stations on land.
Using drifter data, it is possible to obtain a posterior on the underlying
flow. This problem, however, is highly underdetermined. Through controlling
an ocean drifter, we attempt to improve our knowledge of the underlying
flow. We do this by instructing the drifter to explore parts of the flow
currently uncharted, thereby obtaining fresh observations. The efficacy of
a control is determined by its effect on the variance of the posterior
distribution. A smaller variance is interpreted as a better understanding
of the flow. We show a systematic reduction in variance can be achieved by
utilising controls that allow the drifter to navigate new or ‘interesting’
flow structures, a good example of which are eddies.
\end{abstract}

  % \begin{keyword}
  %   uncertainty quantification, Bayesian, inverse problem, data assimilation
  % \end{keyword}

  % \begin{AMS}
  %   62C10, 62F15, 62F30, 62P12, 86A05, 86A22
  % \end{AMS}

  \pagestyle{myheadings}
  \thispagestyle{plain}
  \markboth{D. McDougall and C.~K.~R.~T. Jones}{Decreasing flow uncertainty}

\section{Introduction}
\label{sec:intro}

% Problem statement
% \textcolor{red}{problem statement}

The context of the problem we address in this paper is that of reconstructing a
flow field from Lagrangian observations.  This is an identical twin experiment
in which a true flow field is unknown but from which Lagrangian type
observations are extracted.  It is assumed that little is known about the
functional form of the flow field except that is barotropic, incompressible and
either steady or with simple known time dependence.  Note that, since it is
incompressible and two-dimensional (barotropic), the field is given by a stream
function $\psi (x,y,t)$.  The objective is then to reconstruct an estimate of
this stream function from Lagrangian observations, along with an associated
uncertainty.

The question addressed is whether the uncertainty of the reconstruction can be
reduced by strategic observations using Lagrangian type instruments.  The
measuring devices are assumed to be controllable and their position can be
registered at appropriate time intervals.  Since the control is known, being
prescribed by the operator, it is reasonable to believe that information can be
garnered from the position observations.  The issue is whether we can improve
the information content in these observations by controlling the instruments to
move into specific flow regimes.

% Ocean literature
% \textcolor{red}{ocean literature}

Estimating ocean flows has a long history.  First, a comparison of model
forecast errors in a barotropic open ocean model can be found in
\cite{Robinson1981}, with emphasis on how forecasts are sensitive to boundary
information.  Application of the Kalman filter with Lagrangian observations can
be seen as early as 1982 \cite{Barbieri1982, Miller1986, Parrish1985,
Carter1989}.  For a variational least-squares approach to eddy estimation, the
reader is directed to \cite{Robinson1985}.  A standard mathematical framework
for incorporating Lagrangian observations appeared in 2003
\cite{Kuznetsov2003}.  Finally, \cite{Robel2011} exposes a novel approach to
ocean current observations involving the treatment of sea turtles as Lagrangian
observers.  A good overview of some operational ocean apparatus can be found in
\cite{Rudnick2004}.

% high-level approach
% \textcolor{red}{high-level appraoch}

The underlying philosophy of our approach is that mesoscale ocean flow fields
are dominated by coherent features, such as jets and eddies.  If the instrument
can be controlled to move into and through these structures then the
information gained should be richer in terms of capturing the key properties of
the flow field.

Of course, there is some circularity inherent in this approach; we want to get
the key features of the flow field but need to know them in order to control
the vehicle toward them.  We first take a ``proof of concept'' approach and see
if we follow a simple strategy that we happen to know takes into another eddy,
as opposed to one that doesn’t, then the uncertainty is reduced.  We then
postulate a good way of developing a control based purely on local information.
The idea is to use the known local information of the flow field, from
reconstructing the flow using observations up to a certain point in time, to
form a control that takes the instrument away from the eddy it is currently
stuck in.

% DA intro
% \textcolor{red}{low-level approach/DA intro?}

To obtain the flow reconstruction, one needs to solve a Bayesian inverse
problem \cite{Kalnay2002}.  There are numerous ways to solve Bayesian inverse
probems, with the core methods being Kalman filtering and smoothing
\cite{Kalman1960,Kalman,Sorenson1960,Evensen2006,Houtekamer1998, Anderson2003};
variational methods \cite{Lorenc2000,Bengtsson1975,Lewis1985,
Lorenc1986,LeDimet1986,Talagrand1987,Courtier1994,Lawless2005,Lawless2005a};
particle filtering \cite{Doucet2001,VanLeeuwen2010}; and sampling methods
\cite{Cotter2012,Cotter2009,Cotter2010,Lee2011,Apte2008a,Apte2007,Apte2008,
Herbei2009,Kaipio2000,Mosegaard1995,Roberts1997,Roberts1998,Roberts2001,
Beskos2009,Metropolis1953,Hastings1970,Atchade2005,Atchade2006}.  The resulting
solution to a Bayesian inverse problem is a probability distribution, called
the posterior distribution, over some quanitity of interest from which one can
compute estimates with associated uncertainties.  Bayesian inverse problems
enable well-informed predictions.

% Paper layout
% \textcolor{red}{paper layout}

The paper is organised in the following manner.  The second section sets up the
Bayesian inverse problem and specifies all the assumptions in the prior and
likelihood distributions.  The third section does applies two flow-independent
(na\"ive) controls, a zonal (East-West) control, and a bidirectional
(North-Easterly) control.  These are applied to both the perturbed and
unperturbed flows.  We measure performance of the addition of each by looking
at the posterior variance on the velocity field and show two main results.
When the fluid flow drifter is trapped in a recirculation regime, the magnitude
of the control is the main player in pushing the drifter out of the eddy.  We
show that, for the unperturbed flow, when the control magnitude is large enough
a significant reduction in the posterior variance is achieved.  In the
perturbed flow, we show robustness of the posterior variance with respect to
the perturbation parameter.  More specifically, its structure as a function of
control magnitude is carried over from the time-independent flow model.
Moreover, we observe an additional, and separate, decrease in posterior
variance as a function of control magnitude corresponding to the purely
time-dependent part of the flow.  The fourth section examines the use of an a
posteriori control, a control calculated using information from a previous
Bayesian inversion done with no control present.  Here the control magnitude
corresponds geometrically to the distance between the drifter and a hyperbolic
fixed point of an eddy transport barrier in the flow.  As the control magnitude
increases, the drifter gets closer to a hyperbolic fixed point of the drifter
evolution equation and, for the unperturbed flow, a substantial decrease in
posterior variance is observed.  Hyperbolic fixed points of the drifter
equations join transport barriers in the flow and act as a boundary to
observations.  Observing near these points outweighs the negative effects
produced by polluting the observations with a large control size relative to
the size of the flow.  This gives a novel geometric correspondence between the
control utilised here and the structure of the posterior variance as a function
of control magnitude.  The fifth section concludes the paper.

% Estimating ocean flows from observations of drifter positions is highly
% underdetermined; the path an ocean drifter takes is a minute percentage of the
% observable ocean surface.  Through controlling an ocean drifter, we attempt to
% improve our knowledge of the underlying flow.  We do this by instructing the
% drifter to explore parts of the flow currently uncharted, thereby obtaining
% fresh observations.  The efficacy of a control is determined by its effect on
% the uncertainy of the ocean flow.  A uncertainty is interpreted as a better
% understanding of the flow.  We show a systematic reduction in uncertainty can
% be achieved by utilising controls that allow the drifter to navigate new or
% `interesting' flow structures, a good example of which are eddies.  More
% recently, ocean \textit{gliders} have been designed to scour the Earth's
% oceans, including oceanic structures below the turbulent boundary layer.
% Gliders have the capability to control their roll, pitch and yaw underwater by
% shifting their internal battery to act as a counterweight.  Operationally,
% their objective is to descend into a body of water and take measurements of
% quantities of interest during an ascent, yielding a vertical profile.  Usually
% gliders are equipped with an array of sensors to measure quantities such as
% temperature, concentration of suspended fluorescent particles, salinity, depth,
% and current position.

\section{Setup}
\label{sec:setup}

We begin by prescribing the stream function of the flow field the drifters will
move in.  We will call this flow field the `truth' and later we try to
reconstruct it from noisy observations.  The truth flow we will use is an
explicit solution to the barotropic vorticity equations
\cite{Pierrehumbert1991},
\begin{equation*}
  \psi(x, y, t) = -cy + A \sin(2 \pi k x) \sin(2 \pi y) +
  \varepsilon \psi_1(x, y, t),
  \label{eqn:strfn}
\end{equation*}
on the two dimensional torus $(x, y) \in \mathbb{T}^2$, where the perturbation
we will use is given by
\begin{equation*}
  \psi_1(x, y, t) = \sin (2 \pi x - \pi t) \sin (4 \pi y).
\end{equation*}
The corresponding flow equation is as follows
\begin{equation}
  \diffp{\mathbf{v}}{t} = \varepsilon \partial_t \nabla^{\perp} \psi_1,
  \quad t > 0
  % \diff{\mathbf{x}}{t} &= \mathbf{v}(\mathbf{x}, t), \quad t > 0.
  \label{eqn:model}
\end{equation}
We will explore two cases.  The first case is when $\varepsilon = 0$ and the
underlying flow is steady.  The second case is when $\varepsilon \neq 0$ and
the time-dependent perturbation smears the underlying flow in the
$x$-direction.  Drifters placed in the flow $\mathbf{v}$ will obey
\begin{equation*}
  \diff{\mathbf{x}}{t} = \mathbf{v}(\mathbf{x}, t) + \mathbf{f}(\mathbf{x}, t).
  \label{eqn:control}
\end{equation*}
The function $\mathbf{f}$ is called the \textit{control}, the choice of which
requires explicit diction. We consider two cases of control: a)
flow-independent; and b) a posteriori. Flow-independent controls, are controls
that do not systematically utilise information regarding the underlying flow,
$\mathbf{v}$. A posteriori controls harness information from a previous
Bayesian update. Our soup-to-nuts methodology for assessing the efficacy for
each case of control is as follows. First, drifter dynamics are obtained by
solving
\begin{align}
  \diff{\mathbf{x}}{t} &= \mathbf{v}(\mathbf{x}, t),
  \quad 0 < t \leq t_{K/2} \\
  \diff{\mathbf{x}}{t} &= \mathbf{v}(\mathbf{x}, t) + \mathbf{f}(\mathbf{x}),
  \quad t_{K/2} < t \leq t_K,
  \label{eqn:fullmodel}
\end{align}
where $\mathbf{v}$ solves \eqref{eqn:model}.  Then observations of the drifter
locations $\mathbf{x}$ are collected into an observation vector for both the
controlled and uncontrolled parts
\begin{align}
  \mathbf{y}^1_{k} &= \mathbf{x}(t_k) + \eta_k,
  \quad \eta_k \sim \mathcal{N}(0, \sigma^2 I_2),
  \quad k = 1, \ldots, K, \notag \\
  \leadsto \, \mathbf{y} &= \mathcal{G}(\mathbf{v}_0) + \eta,
  \quad \eta \sim \mathcal{N}(0, \sigma^2 I_{2K}),
  \label{eqn:forward}
\end{align}
where $\mathbf{v}_0$ is the initial condition of the model \eqref{eqn:model}.
The map $\mathcal{G}$ is called the \textit{forward operator} and maps the
object we wish to infer to the space in which observations are taken.

Flow-independent controls $\mathbf{f}$ are independent of $\mathbf{y}^1$. We
will utilise two such controls: a time-independent zonal control
$\mathbf{f}(\zeta, 0)$; and a time-independent bi-directional control
$\mathbf{f}(\zeta, \zeta)$. The a posteriori control we execute is one that
forces drifter paths to be non-transverse to streamlines of the underlying
flow. Namely, $\mathbf{f}(\mathbf{x}) = - \zeta \nabla \mathbb{P}(\psi_0 |
\mathbf{y}^1)$, where $\psi_0(\mathbf{x}) = \psi(\mathbf{x}, 0)$. Our aim is to
understand the effect of the control magnitude $\zeta$ and the resulting
drifter path on the posterior distribution over the initial condition of the
model $\mathbb{P}(\mathbf{v}_0 | \mathbf{y}^1, \mathbf{y}^2)$.

Encompassing our beliefs about how the initial condition, $\mathbf{v}_0$,
should look into a prior probability measure, $\mu_0$, it is possible to
express the posterior distribution in terms of the prior and the data using
Bayes's theorem. Bayes's theorem posed in an infinite dimensional space says
that the posterior probability measure on $\mathbf{v}_0$, $\mu^{\mathbf{y}}$,
is absolutely continuous with respect to the prior proability measure
\cite{Stuart2010}.  Furthermore, the Radon-Nikodym derivative between them is
given by the likelihood distribution of the data,
\begin{equation*}
  \diff{\mu^{\mathbf{y}}}{\mu_0}(\mathbf{v}_0) =
  \frac{1}{Z(\mathbf{y})} \exp \left( \frac{1}{2 \sigma^2} \|
  \mathcal{G}(\mathbf{v}_0) - \mathbf{y} \|^2 \right),
\end{equation*}
where the operator $\mathcal{G}$ is exactly the forward operator as described
in \eqref{eqn:forward} and $Z(\mathbf{y})$ is a normalising constant.  We
utilise a Gaussian prior measure on the flow initial condition, $\mu_0 \sim
\mathcal{N}(0, (- \Delta)^{-\alpha})$. For our purposes, we choose $\alpha = 3$
so that draws from the prior are almost surely in the Sobolev space
$H^1(\mathbb{T}^2)$ \cite{Stuart2010,Bogachev1998}. The posterior is a high
dimensional non-Gaussian distribution requiring careful probing by use of a
suitable numerical method. The reader is referred to \cite{Stuart2010} for a
full and detailed treatment of Bayesian inverse problems on function spaces.

To solve the above Bayesian inverse problem, we use a Markov chain Monte Carlo
(MCMC) method.  MCMC methods are a class of computational techniques for
drawing samples from a unknown target distribution.  Throughout this paper, we
have chosen to sample the posterior using a random walk Metropolis-Hastings
method on function space \cite{Cotter2011,Cotter2012,Bogachev1998}.  Using this
approach, one can draw samples from the posterior distribution, obtaining its
shape exactly.  This is of use when the posterior distribution is not a
Gaussian and cannot be uniquely determined by its first and second moments.
The application of MCMC methods to solve Bayesian inverse problems is
widespread.  For examples of their use, see
\cite{Cotter2009,Cotter2010,Lee2011,Apte2008a,
Apte2007,Apte2008,Herbei2008,Herbei2009,McKeague2005,Michalak2003,Kaipio2000,
Kaipio2007,Mosegaard1995}.  The theory above is all done in an infinite
dimensional setting.  Numerically and operationally, a finite dimensional
approximation is made.  In the case of the Karhunen-Lo\`eve expansion this
approximation is done by truncation.  A choice must be made in where to
truncate, and this choice coincides with a modelling assumption---there are no
frequencies of order larger than the truncation wavenumber.  If it is feasible
that solutions to the inverse problem do in fact admit higher-order
frequencies, it is necessary to rethink this assumption.  Throughout this paper
the data and initial conditions are known and the truncation is chosen to be
much larger than necessary to mitigate the effects of poor modelling
assumptions.  Practically, the true solution to one's problem is unknown.  In
this scenario, care and diligence are necessary in choosing appropriate prior
assumptions.

\section{Results: flow-independent control}
\label{sec:results-crude}

Flow-independent controls concern the influence of an ocean drifter without
using knowledge of the underlying flow. They are constructed in such a way as
to be independent of $\mathbf{y}^1$.

\subsection{Zonal control}
\label{sec:res-hor}

\begin{figure}[htpb]
  \centering
  \subfigure[The norm of the variance decreases as the glider is forced across
  the transport boundary and out of the eddy. The bump occurs as a result of
  the glider exploring a slow part of the flow.]{
    \includegraphics[width=0.45\columnwidth]{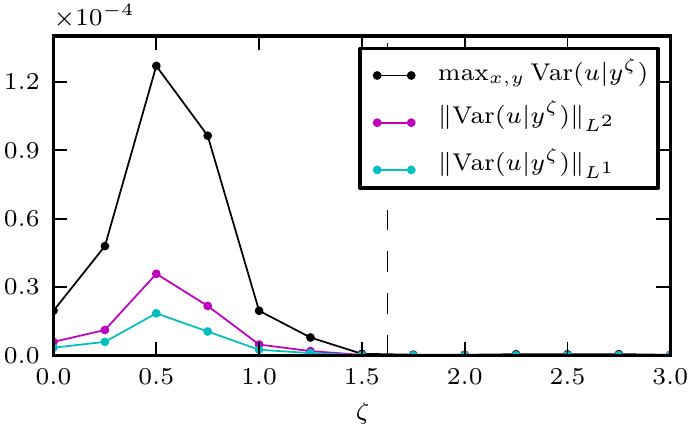}
    \label{fig:ind-xd-norms}
    } \hspace{1em}
  \subfigure[The norm of the variance decreases as the glider is forced across
  the transport boundary and out of the eddy. The second bump appears because
  the glider re-enters a time-dependent eddy.]{
    \includegraphics[width=0.45\columnwidth]{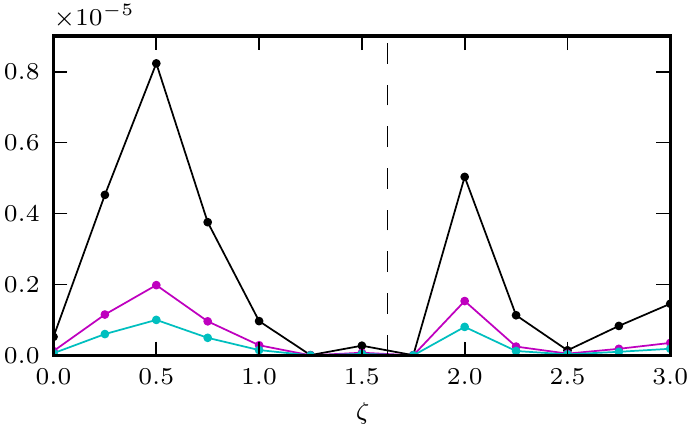}
    \label{fig:dep-xd-norms}
  }
  \caption{Posterior variance as a function of control magnitude, $\zeta$, for
  \subref{fig:ind-xd-norms} the time-independent model; and
  \subref{fig:dep-xd-norms} the time-dependent model.}
  \label{fig:xd-norms}
\end{figure}

Figure~\ref{fig:ind-xd-norms} shows the variance of the horizontal component of
the flow as a function of control magnitude in the max norm, the $L^1$ norm and
$L^2$ norm. The horizontal axis denotes the strength of the control. The
vertical black dotted line corresponds to a critical value for the magnitude.
Values of $\zeta$ less than this correspond to controls not strong enough to
force the drifter out of the eddy. Conversely, values bigger correspond to
controls that push the drifter out of the eddy.

Experiments were done for $\zeta = 0, 0.25, 0.5, \ldots, 2.75, 3$. The case
$\zeta = 1.75$ was the first experiment in which we observed the drifter
leaving the recirculation regime. The black line shows the maximum value of the
variance over the domain $[0, 1] \times [0, 0.5]$. The magenta line and cyan
line show the $L^2$ norm and $L^1$ norm, respectively. The minimum value of the
variance is small enough to be difficult to see on the plot but remains
consistently small, so it has been omitted for clarity reasons. There are some
notable points to make here. Firstly, above the critical value (where the
drifter leaves the eddy) we see that the size of the variance decreases in all
of our chosen norms. We have learned more about the flow around the truth by
forcing the drifter to cross a transport boundary and enter a new flow regime.
Secondly, below the critical region (where the drifter does not leave the eddy)
we see an initial increase in the size of the variance. There are many factors
at play here. We will try to shed some light on them.

\begin{figure}[htpb]
  \centering
  \includegraphics[width=0.45\columnwidth]{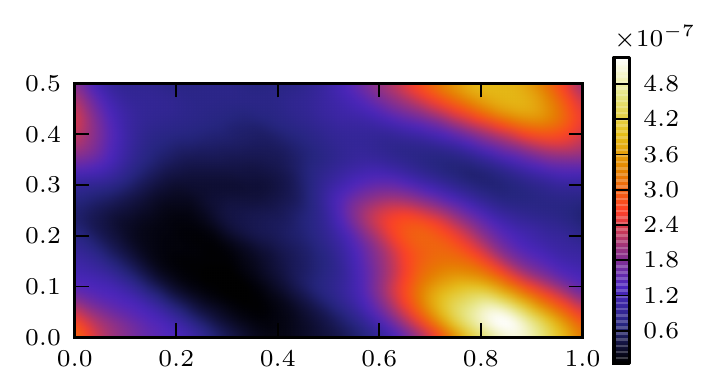}
  \caption{Horizontal component of the posterior variance for the case $\zeta =
  0$. The black area in the lower left corresponds neatly with the region in
  which observations are taken.}
  \label{fig:varu-eg}
\end{figure}

\begin{figure}[htpb]
  \centering
  \includegraphics{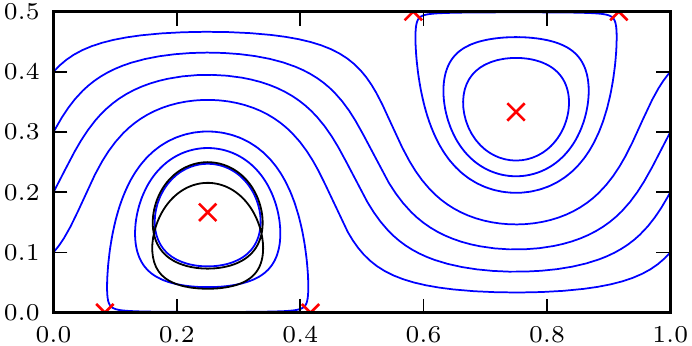}
  \caption{True glider path (black) for some positive $\zeta$ less than the
  critical value. Blue lines are streamlines of the true flow. Red crosses
  are zeros of the flow: fixed points of the passive glider equation.}
  \label{fig:indep-xd-true}
\end{figure}

For small $\zeta$, the controlled and uncontrolled paths along which we take
observations are close. Their closeness and the size of $\sigma^2$ creates a
delicate interplay between whether they are statistically indistinguishable or
not. If they are indistinguishable up to two or three standard deviations, this
could explain the increase and then decrease of the variance below the critical
value.  Secondly, as $\zeta$ increases initially, the controlled path gets
pushed down near the elliptic stagnation point of the flow (see
figure~\ref{fig:indep-xd-true}).  If this region is an area where the flow is
smaller in magnitude than the flow along the uncontrolled path, this is
equivalent to an increase in the magnitude of the control relative to the
underlying flow. This leads to the observations becoming polluted by $f$.

\begin{figure}[htpb]
  \begin{center}
    \includegraphics{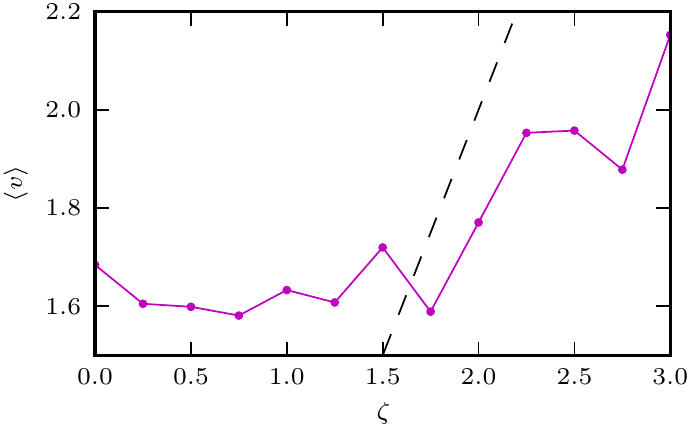}
  \end{center}
  \caption{Mean magitude of the flow along the control path (purple) against
  the size of the control (black dashed line). When the gradient of the flow
  magnitude is large compared with that of the control magnitude, the posterior
  variance is small.}
  \label{fig:xd-rel}
\end{figure}

Exploring this further, we compute the mean magnitude of the flow along the
controlled path of the drifter. More formally, we solve \eqref{eqn:fullmodel}
to obtain a set of points $ \{ zk = z(t_k) \}^{K}_{k=1}$. Then we compute the
mean flow magnitude as follows
\begin{equation}
  \langle v \rangle = \frac{2}{K} \sum_{k=K/2+1}^{K} v(z_k).
  \label{eqn:mean-flow}
\end{equation}
This quantity is computed for each fixed $\zeta$ the result is plotted in
figure~\ref{fig:xd-rel}. The mean flow magnitude is given by the magenta line
in this figure and the black dotted line depicts the flow magnitude. Notice the
first three values of $\zeta$ which the mean flow magnitude decreases in. This
is equivalent to and increase in the magnitude of the control relative to the
magnitude of the underlying flow and so the information gain from taking
observations here decreases. This corresponds nicely with the first three
values of $\zeta$ in figure~\ref{fig:ind-xd-norms} that show an increase in
variance. Notice also that for the other values of $\zeta$ the mean flow
magnitude shows a mostly increasing trend, consistent with a decrease in the
posterior variance.

Note that the region below the critical value correspond to control magnitudes
that are too small to push the glider out of the eddy \textit{in the
unperturbed case} $\varepsilon = 0$. The region above the critical value
corresponds to values of $\zeta$ for which the glider leaves the eddy, this is
also in the unperturbed case. Experiments were done for $\zeta = 0, 0.25, 0.5,
\ldots, 2.75, 3$. In the case $\varepsilon = 0$, the value $\zeta = 1.75$ was
the first experiment in which we observed the glider leaving the recirculation
regime. The black line shows the maximum value of the variance over the domain
$[0, 1] \times [0, 0.5]$. The magenta line and cyan line show the $L^2$ norm
and $L^1$ norm, respectively. There are some notable points to make. Firstly,
below the critical magnitude (where the glider leaves the eddy in the
unperturbed case) we see a sizeable reduction of posterior variance in the
$\max$ norm as the critical magnitude is approached. To establish a connection
in uncertainty quantification between the time independent and time-periodic
case is of great scientific interest and that connection has been made evident
here. Note that as $\zeta$ increases and progresses further into the region
above the critical magnitude, the posterior variance repeats the
increasing/decreasing structure induced by the eddy that we observed in the
region below the critical control magnitude. The new effects introduced into
this region are purely form the time-dependent nature of the moving eddy.  The
reason for their presence is much the same as in the time-independent case;
observations trapped within an eddy regime.
% Notice that for the case $\zeta = 2.0$, the variance is higher and this is
% attributed to the extra loop the true trajectory takes within the eddy. This
% can be seen just north-east of the stationary point depicted by the red
% cross. The true initial condition of the unperturbed PDE is also plotted for
% reference.

We have learned more about the flow around the truth by forcing the glider into
the meandering jet flow regime. The benefits of such a control occur at exactly
the same place as in the time-independent case; as the drifter leaves the eddy
in the unperturbed flow. However, extra care is required when the flow is
time-dependent and the eddy moves. One cannot simply apply the same control
techniques as is evidenced by the extra bump in variance in the region above
the critical magnitude. Of particular use would be extra eddy-tracking
information to construct an a posteriori control to keep the variance small.

\subsection{Bi-directional control}
\label{sec:res-dia}

\begin{figure}[htpb]
  \centering
  \subfigure[The norm of the variance decreases as the glider is forced across
  the transport boundary and out of the eddy.]{
  \includegraphics[width=0.45\columnwidth]{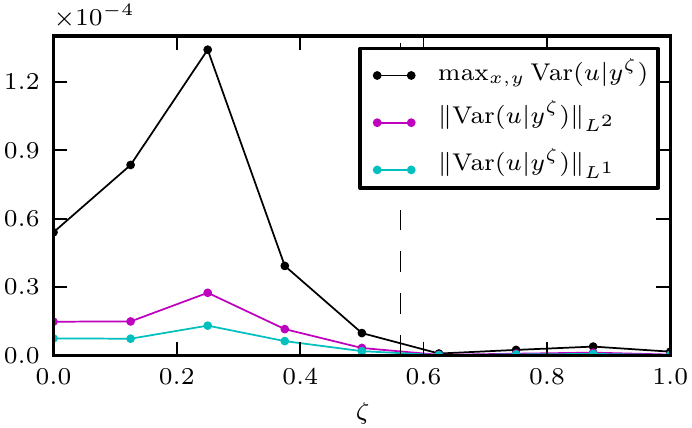}
    \label{fig:ind-xyd-norms}
  } \hspace{1em}
  \subfigure[The norm of the variance decreases as the glider is forced across
  the transport boundary and out of the eddy. The time-dependent part of the
  model pollutes the variance once the glider leaves the eddy.]{
    \includegraphics[width=0.45\columnwidth]{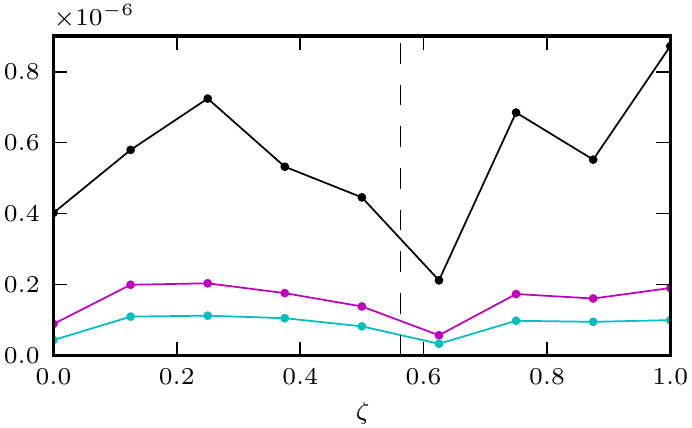}
    \label{fig:dep-xyd-norms}
  }
  \caption{Posterior variance as a function of control magnitude, $\zeta$, for
  \subref{fig:ind-xd-norms} the time-independent model; and
  \subref{fig:dep-xd-norms} the time-dependent model.}
  \label{fig:xyd-norms}
\end{figure}

Now we provide the analogue of figure~\ref{fig:ind-xd-norms} for the bi-directional
forcing function. This is shown in figure~\ref{fig:ind-xyd-norms}. We see
similar behaviour for the variance of the posterior distribution again. Below
the critical magnitude, the values of $\zeta$ for which the drifter is not
forced hard enough to leave the recirculation regime, we see an initial
increase in the size of the posterior variance. Then we observe a decrease in
posterior variance as $\zeta$ approaches a value large enough to push the
drifter out of the eddy regime, the region above the critical value.

\begin{figure}[htpb]
  \centering
  \includegraphics{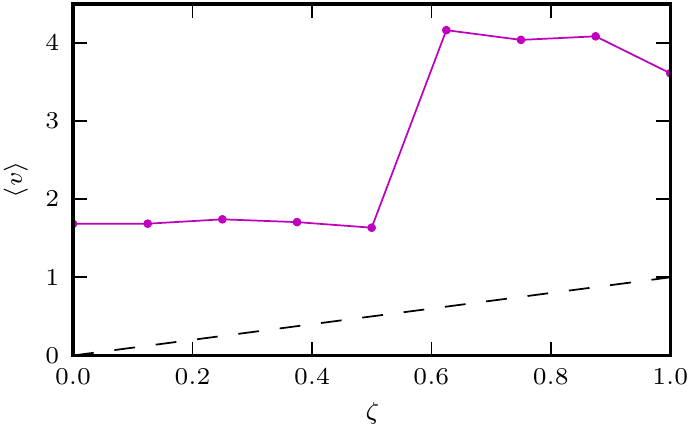}
  \caption{Mean magitude of the flow along the control path (purple) against
  the size of the control (black dashed line). When the gradient of the flow
  magnitude is large compared with that of the control magnitude, the posterior
  variance is small.}
  \label{fig:xyd-rel}
\end{figure}

To explain the initial increase in the posterior variance below the critical
magnitude, we calculate the mean flow magnitude just as in
\eqref{eqn:mean-flow}. This is shown in figure~\ref{fig:xyd-rel}. We see an
initial period where the mean flow along the controlled path remains almost
constant. As a consequence, the magnitude of the forcing increases relative to
the magnitude of the flow.  This pollutes the observations and leads to an
increased posterior variance just as we have observed in the previous section.
We also see the opposite effect; the big jump in flow magnitude at $\zeta =
0.5$ (and consequently when the drifter escapes the gyre) is attested as the
cause of the decrease in posterior variance as we enter the region above the
critical flow magnitude of figure~\ref{fig:ind-xyd-norms}.

The cases of forcing explored thus far are $f(z) = (\zeta, 0)^{\top}$ and $f(z)
= (\zeta, \zeta)^{\top}$. The main results are summarised by referring to
figure~\ref{fig:ind-xd-norms} and figure~\ref{fig:ind-xyd-norms}. In these two
cases, we see strikingly similar structure of the posterior variance as a
function of control magnitude. The initial increase in posterior variance
within the eddy; decreasing posterior variance as the flow path of the drifter
approaches the transport boundary and small posterior variance (compared to the
case $\zeta = 0$) once a new flow regime is being observed. Compare the values
of $\zeta$ for with this behaviour occurs. Notice that the values of $\zeta$ in
figure~\ref{fig:ind-xd-norms} are about three times larger than those in
figure~\ref{fig:ind-xyd-norms}. One factor at play here is the relative
magnitude of the controls in each case. For $\zeta = 1$, the control has
magnitude $1$ in the zonal case, and magnitude $\sqrt{2}$ in the
bi-directional case. Even scaling the results in the bi-directional case by
$\sqrt{2}$, notice that the value of $\zeta$ for which the drifter first leaves
the eddy, is $\zeta = \frac{\sqrt{2}}{2}$ and this is still smaller than $\zeta
= 1.5$ for the $x$-directional case. The final factor affecting the scaling is
the dynamics of the system after the forcing has been applied. Controlling in
only the horizontal direction will require a larger magnitude force to push the
drifter out of the eddy than when forcing in both the $x$ and $y$ directions
simultaneously.

An analogue for figure~\ref{fig:dep-xd-norms} for the new forcing function is
shown in figure~\ref{fig:dep-xyd-norms}. We see similar behaviour for the
variance of the posterior distribution. Again, the region below the critical
magnitude corresponds to values of $\zeta$ that are not big enough to push the
drifter out of the recirculation regime in the \textit{unperturbed} case. Just
as in figure~\ref{fig:dep-xd-norms}, we see the unperturbed eddy affecting the
variance of the posterior distribution on the flow in the classic `bump'
fashion. We observe a reduction in posterior variance as $\zeta$ approaches a
value large enough to push the glider out of the eddy regime (in the case
$\varepsilon = 0$). In the region below the critical magnitude, the
time-dependent flow effects take over and push the variance up. Again, a
connection of uncertainty quantification is made between the time-independent
case and the case where the flow is perturbed by a time-periodic disturbance,
this connection lies entirely within the region below the critical control
magnitude.

\section{Results: a posteriori control}
\label{sec:results-post}

In section~\ref{sec:results-crude} we concluded that crossing a transport
boundary and entering a new flow regime has the desirable effect of reducing
the posterior variance. Crossing into new flow regimes with a stationary flow
can be translated to travelling transversely against the streamlines of the
underlying flow. For the recirculation regime located in the bottom-left area,
particles in the fluid will move in an clockwise fashion.  The gradient of the
stream function will therefore point in towards the fixed point at $z =
\left( 1/4, 1/6 \right)$. The negative gradient of the stream
function points towards the fixed point at $z = \left( 3/4, 1/3
\right)$. Therefore, to escape the recirculation regime we choose,
\begin{equation}
  f(z) = - \zeta \grad_z (\mathbb{E}(\psi | y^1)),
  \label{eqn:gradmeanctrl}
\end{equation}
for the controlled drifter model, where $\psi$ is the stream function of the
flow $v$. The rationale behind this choice is that, if the posterior mean
stream function is a good estimator of the flow, the drifter will be forced
transversely with the stream lines and escape the recirculation regime and
allow us to make observations in a new flow regime.

\begin{figure}[htpb]
  \centering
  \subfigure[The norm of the variance decreases as the glider is forced towards
  the a saddle point in the flow. No clear gain is made otherwise.]{
    \includegraphics[width=0.45\columnwidth]{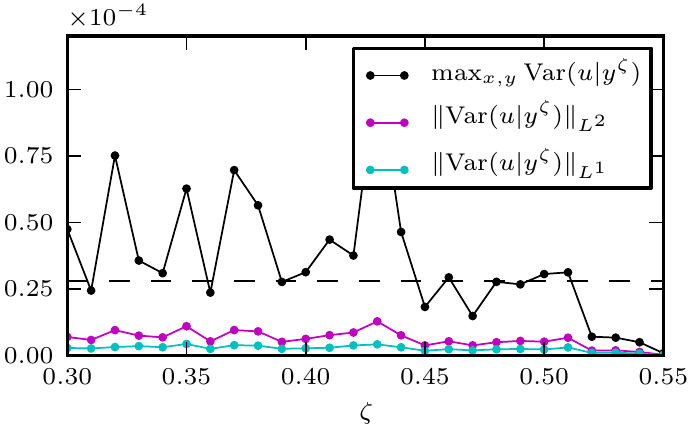}
    \label{fig:ind-grd-norms}
    } \hspace{1em}
  \subfigure[No clear gain is made in the case of the time-dependent model.]{
    \includegraphics[width=0.45\columnwidth]{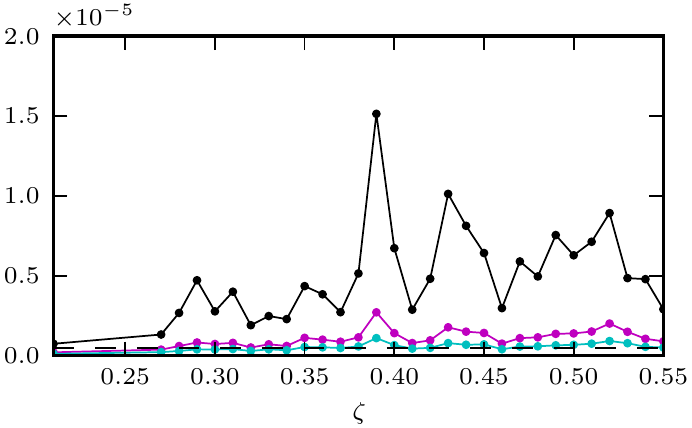}
    \label{fig:dep-grd-norms}
  }
  \caption{Posterior variance as a function of control magnitude, $\zeta$, for
  the a posteriori control in the case of: \subref{fig:ind-xd-norms} a
  time-independent model; and \subref{fig:dep-xd-norms} a time-dependent
  model.}
  \label{fig:grd-norms}
\end{figure}

\begin{figure}[htpb]
  \centering
  \includegraphics[width=0.45\columnwidth]{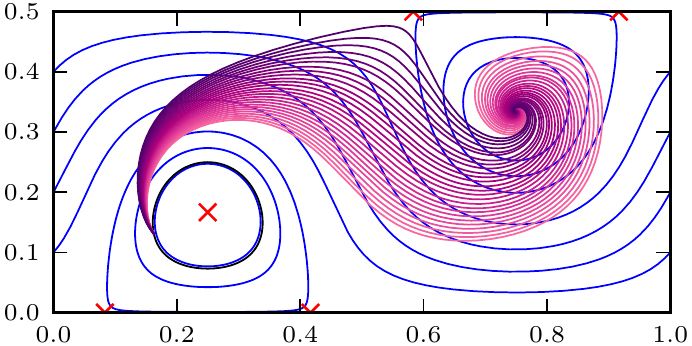}
  \label{fig:ind-grd-true}
  \caption{The true drifter paths for each value of $\zeta$ for the experiments
  shown in figure \ref{fig:ind-grd-norms}. The pink path corresponds to the
  magntidue $\zeta = 0.3$ and the purple path corresponds to $\zeta = 0.55$. The
  posterior variance decays as $\zeta$ approaches $0.55$.}
\end{figure}

Figure~\ref{fig:ind-grd-norms} depicts the variance of the horizontal component
as the strength of the control, $\zeta$, is varied. Note that we do not see the
same behaviour as we do for the two na\"ive controls chosen in
section~\ref{sec:results-crude}.  We see a large band of values of $\zeta$ for
which the posterior variance oscillates, leading to a lack of information gain
in the knowledge of the flow.  From about $\zeta = 0.5$ to $\zeta = 0.55$, we
see a structurally significant reduction in posterior variance where we have a
sustained gain in information about the underlying flow field. This is
attributed to a drifter path that explores an `interesting' part of the flow
where a lot of information can be obtained from observations. To explore the
geometric correspondence between the variance reduction for $\zeta = 0.5$ to
$\zeta = 0.55$, we show figure~\ref{fig:ind-grd-true}. This figure
presents the true path of the drifter for $\zeta = 0.3, \ldots, 0.55$. The
light pink path corresponds to a value of $\zeta = 0.3$ and the purple path
corresponds to $\zeta = 0.55$.  Notice that as $\zeta$ increases, the true path
forms a kink and forms a trajectory close to the zero of the flow at $(x, y) =
(7/12, 1/2)$.  Just as we have seen in section~\ref{sec:results-crude}, we
observe a transient period in the posterior variance until we utilise a control
for which the true path explores new aspects of the flow compared with other
`nearby' controls.  Interestingly, also note that we observe this reduction in
variance despite the true path navigating near a zero of the flow, where we
also satisfy the fact the the size of the control is large in comparison to the
flow. In this case, a logical conclusion here would be that the information
gain from observing near an interesting flow structure heavily outweighs the
information loss in polluting the observations with such a control. The cost of
polluting the observed data can be seen by computing the most structurally
significant reduction in the posterior variance and comparing this with
figure~\ref{fig:ind-xyd-norms}, for example. By `most structurally
significant' we loosely mean the most dramatic reduction that leads to the most
benefit in knowledge of the underlying flow. In this example, this occurs
between $\zeta = 0.52$ and $\zeta = 0.55$, where it is approximately $3 \times
10^{-5}$. In the case of the bi-directional control, where the relative size of
the flow \textit{increases} for the values of $\zeta$ that give a reduction in
variance
% (see figure~\ref{fig:var_vs_z_mu10000_xy} and
% figure~\ref{fig:mean_flow_vs_z_mu10000_xy})
, it occurs between $\zeta = 0.25$
and $\zeta = 0.625$ where it is approximately $1.5 \times 10^{-4}$. This is
about an order of magnitude bigger, crystallising the tradeoff between
polluting the observed data versus exploring `interesting' parts of the flow.
If the posterior mean is a good estimator of the underlying flow, utilising a
control of this nature is beneficial if the drifters navigates close to a
hyperbolic fixed point of the passive drifter model equation.

The first thing to note is that we do not see the same behaviour as we do for
the two na\"ive controls chosen in section~\ref{sec:results-crude}. Nor do we
see similar structures when compared with figure~\ref{fig:ind-grd-norms}. For
each value of $\zeta$, it is the case that the true path navigates to the
time-dependent eddy surrounding the zero of the flow at the point $(x, y) =
\left( 3/4, 1/3 \right)$. The second thing to note is that for all of these
values of control magnitude, the smaller values tend to do better than the
larger ones
% This can be explained by
% figures~\ref{fig:timeper-gm-true0p21}--\ref{fig:timeper-gm-true0p39} which show
% the true drifter paths for the values $\zeta = 0.21, 0.27, 0.39$. In each of
% these plots, the corresponding posterior variance gets larger with $\zeta$. The
% unperturbed initial condition is shown in each plot for reference.
The variance is lower in the cases $\zeta = 0.21$ and $\zeta = 0.27$ because
the true path is navigating towards one of the hyperbolic fixed points of the
eddy. A novel connection is established between the behaviour of these two
controls in both the time-independent case and the time-periodic case.

\begin{figure}[htpb]
  \centering
  \includegraphics[width=0.45\columnwidth]{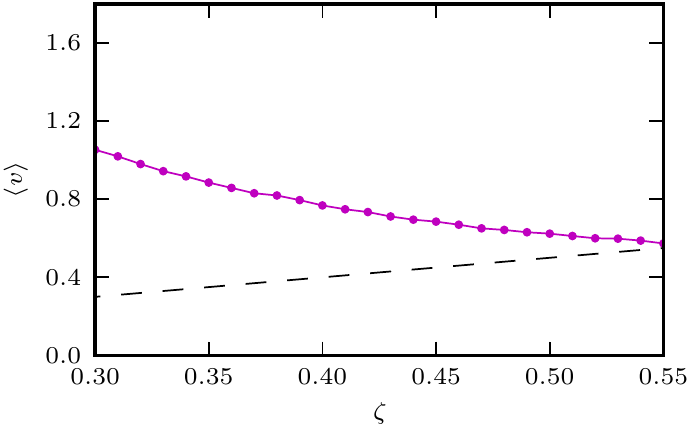}
  \label{fig:ind-grd-rel}
  \caption{Mean magitude of the flow along the control path (purple) against
  the size of the control (black dashed line). Though the gradient of the flow
  magnitude is small compared with that of the control magnitude, the posterior
  variance decreases because the net gain in flow knowledge by observing
  near a saddle point outweighs the net loss by the control polluting
  the observations.}
\end{figure}

\section{Conclusion}
\label{sec:conc}

To summarise, we have measured the performance of two na\"ive control methods,
and one a posteriori control method, both in a time-independent and
time-dependent flow.  We have done so by observing their influence on the
posterior variance in the mean flow direction.  Section~\ref{sec:results-crude}
addresses the na\"ive controls and section~\ref{sec:results-post} the a
posteriori control.  Each control is designed to push ocean drifters into
uncharted flow regimes.  The three cases of control we employ here are a purely
zonal control; a control of equal magnitude in both the $x$ and $y$ directions;
and the gradient of the posterior mean constructed using a posteriori
information from a previous Bayesian update.  In the time-independent flow, we
show a sizeable reduction of the posterior variance in the mean flow direction
for these three cases of control.  We also see that on comparing the posterior
variance for the zonal and bi-directional controls, similar structures arise
when viewed as a function of control magnitude, which dictates when the drifter
leaves the eddy and is the main influence on the posterior information.  In the
case of the a posteriori control in the time-independent flow, the drifter
leaves the eddy for all the values of control magnitude we have chosen.  Here
we observe the variance reduction occurring when the true drifter path
approaches a hyperbolic fixed point on the transport barrier of the eddy in the
upper-right of the domain.  This is evidence that oceanic transport barriers
heavily influence posterior information and sets up a novel geometric
correspondence between the flow structure and the posterior variance.  Using
the na\"ive controls in the time-dependent flow, we show \textit{robustness} of
posterior variance as a function of the perturbation parameter.  When the
control magnitude is such that the drifter leaves the eddy in the
\textit{unperturbed} flow, we see reduction in the posterior variance on the
initial condition for the time-periodic flow.  When employing a time-dependent
a posteriori control, we see no overall net gain in posterior variance over the
uncontrolled case.  For our particular flow and drifter initial condition, it
is the case that the uncontrolled drifter path explores a hyperbolic fixed
point of an eddy in the time-dependent flow more effectively than the
controlled path.  This reiterates the efficacy of control strategies and their
influence on the path along which observations are made.

There are a number of ways in which this work could be generalised in order to
obtain a deeper understanding of the effects controlled ocean drifters have on
flow uncertainty.  For example, (i) the study of non-periodic model dynamics;
(ii) the use of information from the posterior \textit{variance}; (iii) more
elaborate control strategies.  Many other generalisations are also possible.
Non-periodic models are more dynamically consistent with regards to their
approximation of larger ocean models.  We have seen the application of
posterior knowledge in the construction of a control, though only through use
of the mean.  The variance of the underlying flow could be used in a similar
fashion, perhaps to control ocean drifters towards an area of large variance.
This could have a similar affect on the posterior distribution as the method of
controlling a drifter into a new, unexplored flow regime.  Moreover, controls
could be constructed to better reflect reality.  Ocean gliders have a limited
amount of battery power.  Utilising this knowledge in designing a mission plan
to optimise a glider's lifespan certainly has its practical applications.
Controls that minimise the pollution of the observed data is also desirable.
Throughout this paper, we have only used information from one previous Bayesian
update.  Constructing and executing a posteriori control strategies is a
paradigm well suited to that of a Kalman or particle filter; updating the
control every time an analysis step is performed.  This is left for future
discussion.

\section{Acknowledgements}

Author McDougall would like to acknowledge the work of John Hunter
(1968--2012), who led the development of an open-source and freely available
plotting library, matplotlib, capable of producing publication-quality graphics
\cite{Hunter2007}. All the figures in this publication were produced with
matplotlib.

\bibliographystyle{plain}
\bibliography{bibliography}

\end{document}